# Contributions of greenhouse gases and solar activity to global climate change from CMIP6 models simulations

Igor I. Mokhov[1,2] and Dmitry A. Smirnov[1]

[1]A.M. Obukhov Institute of Atmospheric Physics of the Russian Academy of Sciences, 3 Pyzhevsky Per., 119017 Moscow, Russia; mokhov@ifaran.ru

[2]Department of Physics, Lomonosov Moscow State University, Leninskie Gory, 119991 Moscow, Russia

Quantitative estimates of the contributions of the anthropogenic forcing, characterized by changes in the radiative forcing of atmospheric greenhouse gases ($CO_2$, in particular), and solar activity variations to the trends of the global surface temperature on secular temporal horizons are obtained with the aid of autoregressive models from simulations with climate models of the CMIP6 ensemble and from long-term observational data since the 19th century. The results for the simulations with climate models characterized by low, medium and high temperature sensitivity to changes in the $CO_2$ content are compared. It is found, in particular, that the estimates from observation data revealing the determinative contribution of the $CO_2$ content to the global surface temperature trends on half-century and century-long time intervals are most consistent with the estimates from simulations with the climate model with the lowest sensitivity of the global surface temperature to doubling the $CO_2$ atmospheric content.

**Keywords**: current climate change, temperature trends, radiative forcing, atmospheric greenhouse gases, solar activity, CMIP6 ensemble models, long-term data, autoregressive models, directional coupling estimates

## 1. Introduction

Quantitative estimation of the comparative roles of the natural and anthropogenic factors of climate change is one of the key contemporary problems. According to (Bindoff et al., 2013), with probability greater than 90 % more than half of the global surface temperature (GST) increase since the middle of the 20th century relates to the anthropogenic increase of the atmospheric content of the greenhouse gases (GHG), see also (Climate Change, 2021; Мохов, 2022).

Considerable influence of the GHG atmospheric content increase on the current GST increase is revealed from observation data under the account of various natural factors, including solar and volcanic activity and different climate variability modes, in many studies (Climate Change, 2021; Santer et al., 2001; Allen et al., 2006; Kaufmann et al., 2006; Lockwood, 2008; Foster and Rahmstorf, 2011; Kopp and Lean, 2011; Kaufmann et al., 2011; Loehle and Scafetta, 2011; Gruza and Ran'kova, 2012; Zhou and Tung, 2013; Stern and Kaufmann, 2014; Stolpe et al., 2017; Mokhov and Smirnov, 2018a,b; Kajtar et a.l, 2019; McBride et al., 2021). In many works the contribution of GHGs and other factors to temperature changes was statistically analyzed using time series without additional physical assumptions (i.e. any climate models), see e.g. (Tol and de Vos, 1993; Kaufmann and Stern, 1997; Lean and Rind, 2008; Mokhov and Smirnov, 2009; Smirnov and Mokhov, 2009; Attanasio and Triacca, 2011;

Kodra et al., 2011; Mokhov et al., 2012; Triacca et al., 2013; Mokhov and Smirnov, 2016; Stips et al., 2016; Mokhov and Smirnov, 2022).

As for an expected climate change, its adequate estimate is achievable only with the aid of climate models. Still, variability of model-based estimates is quite large not only due to uncertainties in possible anthropogenic impact scenarios, but also due to different sensitivities of model climate characteristics to external forcings. In particular, the sensitivity of the GST to the $CO_2$ atmospheric content is quite diverse. According to (Scafetta, 2022; Scafetta, 2023), one can distinguish CMIP6 climate models with high (from 4.5 K up to 5.6 K), medium (from 3.0 to 4.5 K) and low (from 1.8 to 3.0 K) sensitivity of the GST to doubling the $CO_2$ atmospheric content.

This work presents estimates of the contributions of the $CO_2$ atmospheric content and insolation variations to the GST trends on various temporal horizons from simulations with the CMIP6 climate models in comparison with empirical estimates from the instrumental observation data.

## 2. Data and method

To estimate contributions of the $CO_2$ atmospheric content and insolation variations to the GST trends, we have used simulations with 7 different CMIP6 models under the "historical" scenario of the natural and anthropogenic influences on the climate system over the period 1850 – 2014: INM-CM4-8, INM-CM5-0, MIROC-ES2L-f2, BCC-CSM2-MR, CNRM-CM6-1-HR-f2, CNRM-ESM2-1-f2 and CNRM-CM6-1-f2. The choice of the models from the CMIP6 ensemble is determined by different sensitivities of their temperature regimes to external forcings, in particular, to the $CO_2$ atmospheric content variations. Table 1 presents the sensitivity parameter for the selected CMIP6 climate models (Scafetta, 2023) including its low (L), medium (M) and high (H) values. This parameter $I_{ECS}$ denotes the equilibrium GST change under doubling the $CO_2$ atmospheric content.

Table 1. Parameter $I_{ECS}$ indicates low (L), medium (M) or high (H) sensitivity of the GST to doubling the $CO_2$ atmospheric content according to simulations with CMIP6 climate models, see (Scafetta, 2023).

| | CMIP6 models | $I_{ECS}$, K |
|---|---|---|
| L | INM-CM4-8 | 1.8 |
| L | INM-CM5-0 | 1.9 |
| L | MIROC-ES2L-f2 | 2.7 |
| M | BCC-CSM2-MR | 3.0 |
| M | CNRM-CM6-1-HR-f2 | 4.3 |
| H | CNRM-ESM2-1-f2 | 4.8 |
| H | CNRM-CM6-1-f2 | 4.8 |

The estimates from the simulated model-based data are compared here with the corresponding estimates from the empirical HadCRUT5 data (https://www.metoffice.gov.uk/hadobs/hadcrut5/) for the annual GST over the time interval 1880 – 2012 (Mokhov and Smirnov, 2018a,b, 2023). Contributions of the $CO_2$ atmospheric content and insolation variations (on the basis of their values used in simulations with the CMIP6 models, https://esgf-data.dkrz.de/projects/cmip6-dkrz/) to the GST trends on the time intervals of the length ranging from 10 to 130 years are estimated. As the GST values, we have used both the "surface temperature" (the variable ts in the CMIP6 data) and the "surface air temperature" (the variable tas in the CMIP6 data).

Contributions of the two factors under consideration to the GST trends are estimated with the aid of empirical trivariate autoregressive (AR) models for the GST denoted here as $T$ [K]:

$$T_n = a_0 + a_1 T_{n-1} + a_2 I_{CO_2, n-1} + a_3 I_{S, n-1} + \xi_n \ , \tag{1}$$

where the discrete time $n$ means the calendar year, $I_{CO_2}$ [W/m$^2$] is the $CO_2$ radiative forcing, $I_S$ [W/m$^2$] is the time-dependent "solar constant". Along with the analysis of the trends in moving windows of a fixed length (from 10 to 130 years), we have performed the analysis of the trends in time windows with the fixed endpoint of 2012 and the starting point sliding from 1880 to 2002. We have estimated coupling coefficients and the level (variance) of the noise $\xi$ which reflects various processes unaccounted for in the AR model. In order to estimate the contribution of a given factor (either the $CO_2$ radiative forcing or the solar activity variations) to the GST trends, we first subtract the GST time realization obtained as an output of the AR model under the hypothetic condition (that the factor under consideration is constant and equals its value observed in 1880) from the original GST time series which has been used to fit the AR model. Second, the contribution of the factor under consideration to the linear trend of the GST on a time interval $[L_{start}, L_{end}]$ of the length $L = L_{end} - L_{start}$ is estimated as the linear trend of the difference between the above two time realizations of the GST. The value of the linear trend (i.e. the angular coefficient of the straight line approximating the temporal profile of the GST in a given time window) is determined via the ordinary least-squares technique. Errors in the estimates of the AR coefficients, linear trends and contributions to the trends are determined through the usual multiple regression formalism. In particular, estimates of statistical significance levels for the conclusions about nonzero contributions to the GST trends are obtained as the value of the inverse cumulative distribution function of the standard Gaussian law whose argument is the coupling coefficient estimate divided by the estimate of its standard error. In more detail, the method is given in (Mokhov and Smirnov, 2018a,b, 2023), see also (Smirnov and Mokhov, 2009; Mukhin et al, 2021).

## 3. Results

Figure 1 presents the GST anomalies (relative to the base period 1961 – 1990) since 1880 according to the HadCRUT5 data and simulations with 7 climate models of the CMIP6 ensemble under the scenario "historical". Different models exhibit essentially different long-term variations and inter-annual variability of the GST.

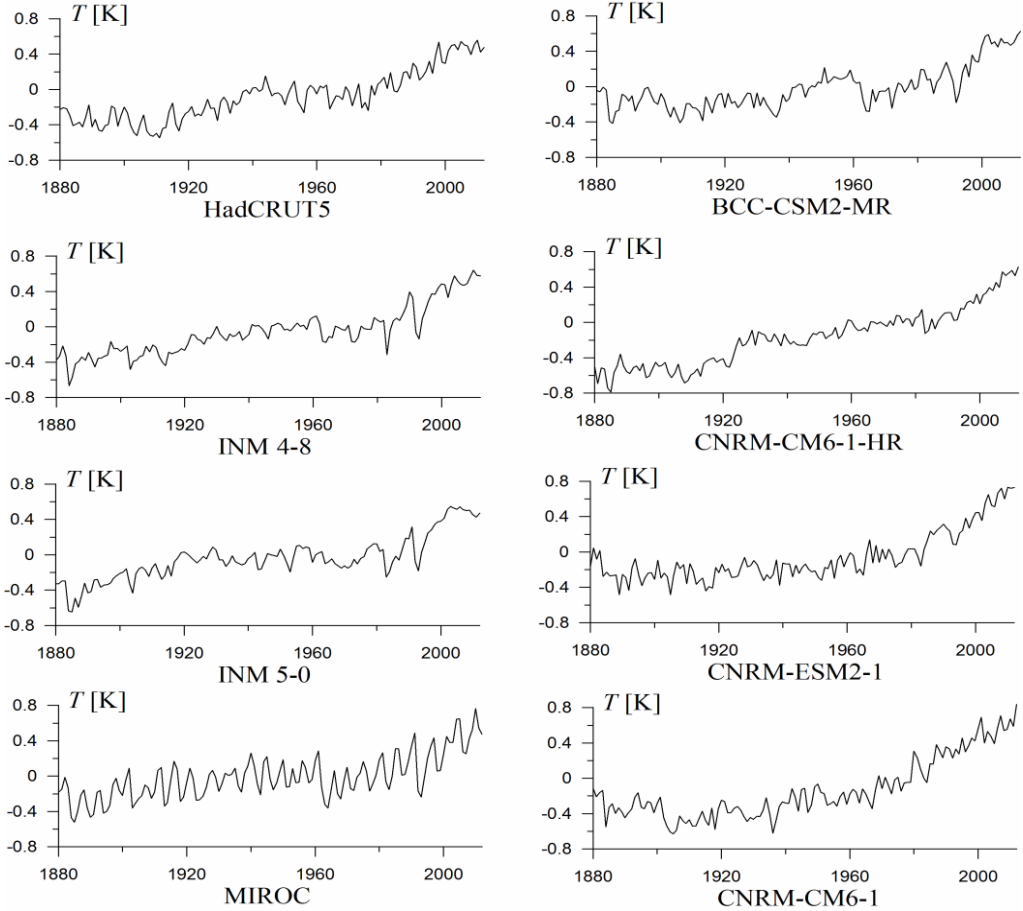

Figure 1. Inter-annual variations of the GST anomalies since 1880 according to the HadCRUT5 data and simulations with different CMIP6 models under the scenario "historical".

Table 2 shows the coefficients of the AR models (1) obtained from the CMIP6 model simulations in comparison with the corresponding estimates from the HadCRUT5 data since 1880. Thus, the estimates of $a_2$ (coefficient of coupling between the GST and the GHG radiative forcing) obtained from the model simulations range from 0.06 $K \cdot m^2/W$ to 0.17 $K \cdot m^2/W$ with the semi-width of the 95% confidence interval ranging from 0.03 $K \cdot m^2/W$ to 0.05 $K \cdot m^2/W$. The corresponding estimates from the HadCRUT5 data are $0.13 \pm 0.04$ $K \cdot m^2/W$ and so belong to the range of the estimates obtained from the CMIP6 model simulations. The largest values of $a_2$ correspond to the CMIP6 models with the largest sensitivity of the GST to doubling the $CO_2$ atmospheric content $I_{ECS}$. The estimates of $a_3$ (coefficient of coupling between the GST and the insolation variations) obtained both from the HadCRUT5 data (equal to $0.03 \pm 0.05 \, \text{K} \cdot \text{m}^2/\text{W}$) and from the CMIP6 model simulations are statistically insignificant. The estimates from the HadCRUT5 data are most consistent with those from simulations with the CMIP6 models with medium sensitivity. The estimates of the AR coefficient $a_1$ (which characterizes "inertia" of the temperature regime, dimensionless quantity close to unity for a large relaxation time and to zero for a small one) for the CMIP models range from 0.42 to 0.76 with the semi-width of the 95% confidence interval ranging from 0.12 to 0.16. The values closest to the estimates from the HadCRUT data ($0.54 \pm 0.15$) are obtained for the CMIP6 models with high sensitivity $I_{ECS}$.

Table 2. Coefficients $a_1$ (dimensionless), $a_2$ [$\text{K} \cdot \text{m}^2/\text{W}$] and $a_3$ [$\text{K} \cdot \text{m}^2/\text{W}$] of the AR model (1) obtained from the HadCRUT data since 1880 and from the CMIP6 model simulations with low (L), medium (M) and high (H) sensitivity of the GST to doubling the $CO_2$ atmospheric content.

|  |  |  | $a_1(\pm 2\sigma_{a_1})$ | $a_2(\pm 2\sigma_{a_2})$ | $a_3(\pm 2\sigma_{a_3})$ |
|---|---|---|---|---|---|
| data |  | HadCRUT5 | **0.54 (±0.15)** | **0.13 (±0.04)** | **0.03 (±0.05)** |
| models | L | INM 4-8 | 0.64 (±0.13) | 0.10 (**±0.04**) | 0.01 (**±0.05**) |
|  |  | INM 5-0 | 0.76 (±0.12) | 0.06 (±0.03) | -0.00 (±0.04) |
|  |  | MIROC-ES2L-f2 | 0.42 (±0.16) | **0.12** (±0.05) | 0.01 (±0.08) |
|  | M | BCC-CSM2-MR | 0.65 (±0.13) | 0.08 (±0.03) | **0.02 (±0.05)** |
|  |  | CNRM-CM6-1-HR-f2 | 0.73 (±0.12) | 0.09 (**±0.04**) | 0.02 (±0.04) |
|  | H | CNRM-ESM2-1-f2 | 0.49 (**±0.15**) | 0.15 (±0.05) | -0.03 (±0.06) |
|  |  | CNRM-CM6-1-f2 | **0.55** (±0.14) | 0.17 (±0.05) | **0.02** (±0.06) |

Table 3. The estimates of the $CO_2$ atmospheric content contribution $C_{CO_2}$ [K/decade] and relative contribution $C_{CO_2}/\alpha_T$ (dimensionless) to the GST trends $\alpha_T$ [K/decade] on the temporal horizons of the length $L$ of 50 and 130 years.

|  |  |  | $L = 50$ years | | $L = 130$ years | |
|---|---|---|---|---|---|---|
|  |  |  | $C_{CO_2}$ | $C_{CO_2}/\alpha_T$ | $C_{CO_2}$ | $C_{CO_2}/\alpha_T$ |
| data |  | HadCRUT5 | **0.13** | **0.90** | **0.06** | **0.94** |
| models | L | INM-CM4-8 | **0.13** | **0.84** | **0.06** | **0.97** |
|  |  | INM-CM5-0 | 0.11 | 0.80 | 0.05 | 0.99 |
|  |  | MIROC-ES2L-f2 | 0.10 | 0.67 | 0.05 | 1.01 |
|  | M | BCC-CSM2-MR | **0.11** | **0.71** | **0.05** | **1.04** |
|  |  | CNRM-CM6-1-HR-f2 | **0.16** | **1.26** | **0.07** | **0.93** |
|  | H | CNRM-ESM2-1-f2 | 0.14 | 0.79 | 0.07 | 1.12 |
|  |  | CNRM-CM6-1-f2 | 0.17 | 0.90 | 0.08 | 1.05 |

Table 3 presents the estimates of the $CO_2$ contribution $C_{CO_2}$ and relative contribution $C_{CO_2}/\alpha_T$ to the GST trend $\alpha_T$ on the temporal horizons of the length $L$ of 50 and 130 years obtained from the HadCRUT5 data and the CMIP6 model simulations. One can see the closest correspondence between the estimate from the HadCRUT5 data (revealing the determinative contribution of the $CO_2$ atmospheric content to the GST trend on the half-century and century-long horizons) and the estimate from simulations with the climate model INM-CM4-8 with the lowest sensitivity of the GST to doubling the $CO_2$ atmospheric content. Relative errors of the obtained $C_{CO_2}$ estimates are equal to the relative errors of the estimates of the corresponding AR coefficient $a_2$ in (1).

Figure 2 shows the estimates of the $CO_2$ atmospheric content contributions to the GST trends on various time intervals from the HadCRUT5 data and the CMIP6 model simulations. The estimates are shown versus the endpoint $L_{end}$ of the 15-year time windows (Fig. 2a) and the starting point $L_{start}$ of the time windows with the fixed endpoint $L_{end} = 2012$ (Fig. 2b). The estimates for the 130-year time window are statistically significant (at $p$-level much less than 0.05, see Table 1) and range from 0.05 K/decade for the model MIROC to 0.08 K/decade for the model CNRM-CM6-1, while the estimate from the HadCRUT5 data equals 0.06 K/decade (Fig. 2b). The relative $CO_2$ atmospheric content contribution to the GST trend on the 130-year interval is estimated to be in the range from 0.93 to 1.12, while its estimate from the HadCRUT5 data equals 0.94 (Table 3).

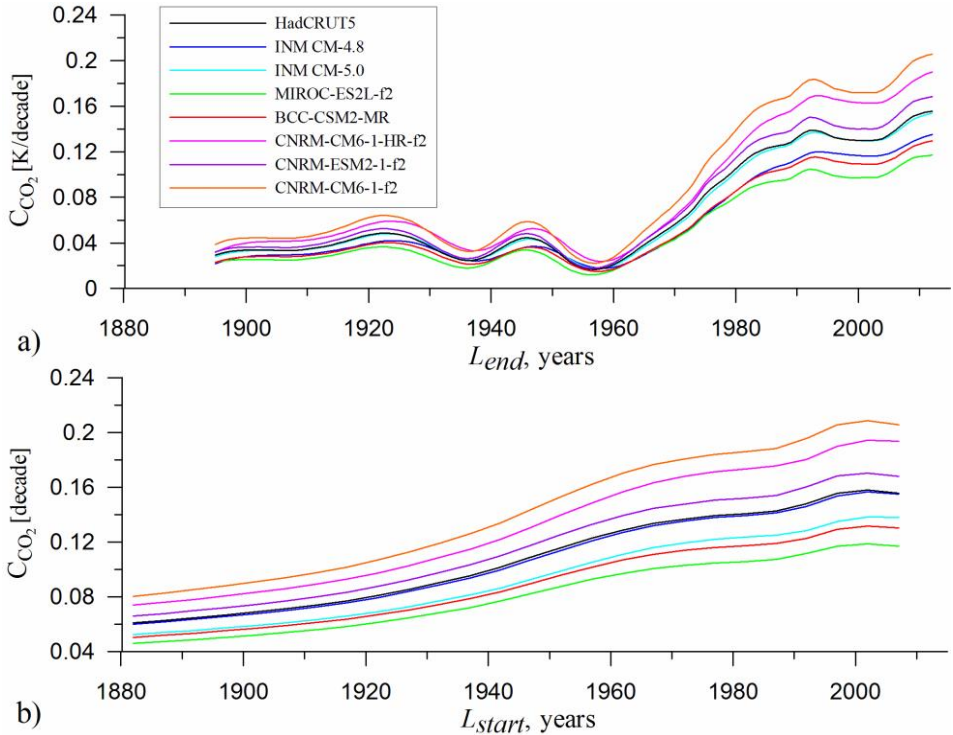

Figure 2. Estimates of the $CO_2$ atmospheric content contributions to the GST trends from the CMIP6 model simulations and the HadCRUT5 data: (a) versus the endpoint $L_{end}$ of the 15-year intervals, (b) versus the starting point $L_{start}$ of the intervals with the fixed endpoint $L_{end} = 2012$.

Table 4 shows the estimates of the insolation variations contribution $C_S$ and its relative contribution $C_S/\alpha_T$ to the GST trends $\alpha_T$ on the time intervals of the length $L$ of 50 and 130 years from the HadCRUT data and the CMIP6 model simulations. The estimates of $C_S$ are statistically insignificant even at $p$-level 0.1. The relative contribution $C_S/\alpha_T$ estimated from the HadCRUT5 data equals 5 % for the 130-year time window and 2 % for the 50-year time window. These estimates correspond best to the estimates for the climate models with the medium sensitivity $I_{ECS}$.

Table 4. The estimates of the insolation variation contribution $C_S$ [K/decade] and its relative contribution $C_S/\alpha_T$ (dimensionless) to the GST trends $\alpha_T$ [K/decade] on the temporal horizons of the length $L$ of 50 and 130 years.

|  |  |  | $L = 50$ years | | $L = 130$ years | |
| --- | --- | --- | --- | --- | --- | --- |
|  |  |  | $C_S$ | $C_S/\alpha_T$ | $C_S$ | $C_S/\alpha_T$ |
| data |  | HadCRUT5 | **-0.003** | **-0.02** | **0.003** | **0.05** |
| models | L | INM-CM4-8 | -0.001 | -0.01 | 0.002 | 0.03 |
|  | L | INM-CM5-0 | 0.0004 | 0.003 | -0.0006 | -0.01 |
|  | L | MIROC-ES2L-f2 | -0.0003 | -0.002 | 0.0004 | 0.01 |
|  | M | BCC-CSM2-MR | **-0.002** | **-0.02** | **0.003** | **0.06** |
|  | M | CNRM-CM6-1-HR-f2 | **-0.002** | **-0.02** | **0.003** | **0.04** |
|  | H | CNRM-ESM2-1-f2 | 0.002 | 0.01 | -0.003 | -0.04 |
|  | H | CNRM-CM6-1-f2 | -0.002 | -0.01 | 0.002 | 0.02 |

Similar results are obtained for other time windows with somewhat greater contributions of the $CO_2$ atmospheric content to the GST trends on shorter time intervals, i.e. during the last decades. We also note that the contribution estimation results from the model variables ts and tas serving as the GST representatives are almost indistinguishable.

## 4. Conclusions

Quantitative estimates of the contributions of the anthropogenic influence, characterized by the changes of the $CO_2$ radiative forcing, and solar activity variations to the GST trends on century-long temporal horizons are obtained from CMIP6 climate model simulations and the HadCRUT5 data since the middle of the 19 century with the aid of AR empirical models. For all climate models and the observation data under consideration, the estimated contributions of the $CO_2$ atmospheric content to the GST trends are statistically significant and relatively quite large, while the estimated contributions of the insolation variations are statistically insignificant.

The results for the climate models with low, medium and high sensitivity of the GST to doubling the $CO_2$ atmospheric content are compared. In particular, it is obtained that the estimates from the HadCRUT5 data, revealing the determinative contribution of the $CO_2$ atmospheric content to the GST trends on half-century and century-long temporal horizons, are most close to the estimates from simulations with the climate model INM-CM4-8 which exhibits the lowest sensitivity $I_{ECS}$ among the 7 models under consideration. It is known that sensitivity of the model climate to external influences essentially depends on the parameterization of cloudiness (Marchuk et al., 1986), see also (Volodin, 2021; Golitsyn and Mokhov, 1978; Mokhov, 1979; Mokhov et al., 1994; Mokhov, 1991). In (Volodin, 2021), a strong dependence of the equilibrium sensitivity of the climate model INM-CM4-8 on the parameterization of cloudiness has been shown. It has been noted that the replacement of the diagnostic parameterization (with linear dependence of the percentage of a model box on the relative humidity of the atmosphere) by the prognostic parameterization (Tiedke, 1993) increases the equilibrium sensitivity of the GST to doubling the $CO_2$ atmospheric content more than twice. Under the usage of the diagnostic parameterization, warming in the model is accompanied by the low-level cloudiness related to the inversion on the top boundary of the atmospheric boundary layer. Under the usage of the prognostic parameterization (Tiedke, 1993), cloudiness under warming decreases at all atmospheric levels. Switching off the mechanism, which induces the decrease of the deep convection cloudiness under warming, increases the equilibrium sensitivity even more. Switching off the mechanism, which induces the decrease of the atmospheric boundary layer cloudiness under warming, provides conditions in favor of the decrease of the model sensitivity.

The work is carried out under the financial support of the RSF project no. 24-17-00211 using the results obtained within the framework of the RSF project no. 23-47-00104.